\newif\ifhasbib
\newif\ifsubmit
\newif\ifastroph

\hasbibtrue

\submittrue

\astrophtrue 

\ifsubmit
	\ifastroph
		\documentclass[twocolumn]{aastex631}
	\else
		\documentclass[twocolumn,linenumbers]{aastex631}
	\fi
\else
	\documentclass[twocolumn]{aastex631}
\fi

\shorttitle{Pebbles}
\shortauthors{Mulders et al.}

\usepackage{natbib,graphicx,amsmath,xspace,footnote}
\usepackage[T1]{fontenc}

\usepackage{xcolor, fontawesome, microtype}
\definecolor{twitterblue}{RGB}{64,153,255}
\definecolor{linkcolor}{rgb}{0.1216,0.4667,0.7059}

\newcommand{\twitter}[1]{\href{https://twitter.com/#1}{\textcolor{twitterblue}{\faTwitter}\,\tt \textcolor{twitterblue}{@#1}}}

\definecolor{darkblueish}{RGB}{59, 64, 139}

\begin{document}

\title{Why do M dwarfs have more transiting planets?}

\correspondingauthor{Gijs D. Mulders}
\email{gijs.mulders@uai.cl}


\author[0000-0002-1078-9493]{Gijs D. Mulders}
\affil{Facultad de Ingenier\'ia y Ciencias, Universidad Adolfo Ib\'a\~nez, Av.\ Diagonal las Torres 2640, Pe\~nalol\'en, Santiago, Chile \twitter{GijsMulders}}
\affil{Millennium Institute for Astrophysics, Chile}
\affil{Alien Earths Team, NASA Nexus for Exoplanet System Science}

\author[0000-0002-9128-0305]{Joanna Dr{\k{a}}{\.z}kowska} 
\affil{University Observatory, Faculty of Physics, Ludwig-Maximilians-Universit\"at M\"unchen, Scheinerstr. 1, 81679 Munich, Germany}

\author[0000-0003-2458-9756]{Nienke van der Marel}
\affil{Physics \& Astronomy Department, 
University of Victoria, 
3800 Finnerty Road, 
Victoria, BC, V8P 5C2, 
Canada}

\author[0000-0002-0093-065X]{Fred J. Ciesla}
\affil{Department of the Geophysical Sciences, The University of Chicago, 5734 South Ellis Avenue, Chicago, IL 60637}
\affil{Alien Earths Team, NASA Nexus for Exoplanet System Science}

\author[0000-0001-7962-1683]{Ilaria Pascucci}
\affil{Lunar and Planetary Laboratory, The University of Arizona, Tucson, AZ 85721, USA}
\affil{Alien Earths Team, NASA Nexus for Exoplanet System Science}
%


\begin{abstract} 
We propose a planet formation scenario to explain the elevated occurrence rates of transiting planets around M dwarfs compared to sun-like stars discovered by \textit{Kepler}. 
We use a pebble drift and accretion model to simulate the growth of planet cores inside and outside of the snow line. A smaller pebble size interior to the snow line delays the growth of super-Earths, allowing giant planet cores in the outer disk to form first. When those giant planets reach pebble isolation mass they cut off the flow of pebbles to the inner disk and prevent the formation of close-in super-Earths. We apply this model to stars with masses between 0.1 and 2 $M_\odot$ and for a range of initial disk masses. 
We find that the masses of hot super-Earths and of cold giant planets are anti-correlated.  
The fraction of our simulations that form hot super-Earths is higher around lower-mass stars and matches the exoplanet occurrence rates from \textit{Kepler}.
The fraction of simulations forming cold giant planets is consistent with the stellar mass dependence from radial velocity surveys.
A key testable prediction of the pebble accretion hypothesis is that the occurrence rates of super-Earths should decrease again for M dwarfs near the sub-stellar boundary like Trappist-1.
\end{abstract}



\section{Introduction}
Exoplanets provide important insights into the processes that operate during planet formation.
The discovery of 
super-Earths --- planets with masses and radii between that of Earth and Neptune and orbiting within 1 au of their host stars --- have spawned several new hypotheses about how planets are assembled given that these planets are absent in our own solar system.
Most of these scenarios involve either the radial drift of pebble-sized solids or the migration of earth-sized proto-planets -- in a contrast with formation models the solar system terrestrial planets which are mostly ``in situ''. 
As exoplanet surveys continue to discover planets and the overall view of exoplanet demographics becomes more clear, these planet formation hypotheses need to be evaluated against the known exoplanet populations.

One observed trend in exoplanets demographics that has so far eluded a straightforward explanation is the elevated occurrence rate of super-Earths around low-mass stars (see, e.g. \citealt{2018arXiv180500023M} for a review). The occurrence rates of super-Earths observed with \textit{Kepler} increase from F stars to M dwarfs by a factor $~3$ \citep{2012ApJS..201...15H,2015ApJ...798..112M}, a result that has been recently confirmed by radial velocity surveys \citep{sabotta:2021ij}.
These higher occurrence rates have been show to correspond to a higher fraction of stars with planetary systems \citep{2020AJ....159..164Y,2021AJ....161...16H}.


This trend defies the scaling relations between host star mass and giant planet occurrence \citep[e.g.][]{2010PASP..122..905J,2018ApJ...860..109G,2021arXiv210511584F}, between host star mass and protoplanetary disk dust mass \citep[e.g.][]{2016ApJ...831..125P}, and is typically not recovered in planet population synthesis models \citep[e.g.][]{2021arXiv210504596B}. 
Thus, it likely points to an incomplete description of the formation of super-Earths in conventional planet formation models. 

The pebble accretion hypothesis involves the direct accretion of centimeter-sized pebbles onto forming proto-planets within a gaseous disk \citep[e.g.][]{2010A&A...520A..43O}. It was initially proposed to accelerate the growth of Jupiters core \citep{2012A&A...544A..32L,2015Natur.524..322L} and it can also contribute to the growth of super-Earths if the radial flux of pebbles drifting inward to the inner disk is high \citep[e.g.][]{2019A&A...627A..83L}. Pebble accretion is thus a viable way of forming super-Earths around M dwarfs \citep{2019A&A...632A...7L,2019A&A...627A.149S}. 
However, the pebble flux into the inner disk scales positively with stellar mass because it is proportional to the total dust mass, and thus super-Earth formation is expected to be more efficient around sun-like stars, contrary to what is observed.
Therefore an additional mechanism is needed to suppress the formation of super-Earths around sun-like stars that does not operate around M dwarfs to be consistent with exoplanet occurrence rates. 

A mechanism to reduce the pebble flux is filtering by giant planets, which open gaps in the disk and blocks the inward radial flow of pebbles, preventing the growth of super-Earths \citep[e.g.][]{2019A&A...627A..83L}. 
Because giant planets are more common around more massive stars, this suppression of super-Earth formation could be less effective around low-mass stars.
Recently, \cite{2021arXiv210406838V} showed that this hypothesis is consistent with 
the higher frequency of disk gaps at tens of au around higher mass stars observed with ALMA, 
and that drift-dominated disks without those gaps are more frequent around M dwarfs, those stars with high occurrence rates of super-Earths.

In this paper, 
we explore the conditions under which an anti-correlation between super-Earth occurrence and stellar mass arises within a pebble accretion framework. 
In Section \ref{s:model} we describe our model framework and the conditions under which the growth of super-Earths is quenched by giant planet formation, highlighting the role of snow lines as a crucial additional ingredient to delay the formation of hot super-Earths. In Section \ref{s:para} we quantify with a parameter study in disk and stellar mass how planet occurrence rates would vary with stellar mass, and in Section \ref{s:obs} we compare these predictions to observations of radial velocity and transit surveys. We conclude by discussing essential tests and possible challenges for this hypothesis: a predicted decrease in planet occurrence rate towards the stellar/brown dwarf boundary and the strength of the observed super-Earth giant planet correlation.

\begin{figure}
    \centering
    \ifsubmit
    \includegraphics[width=\linewidth]{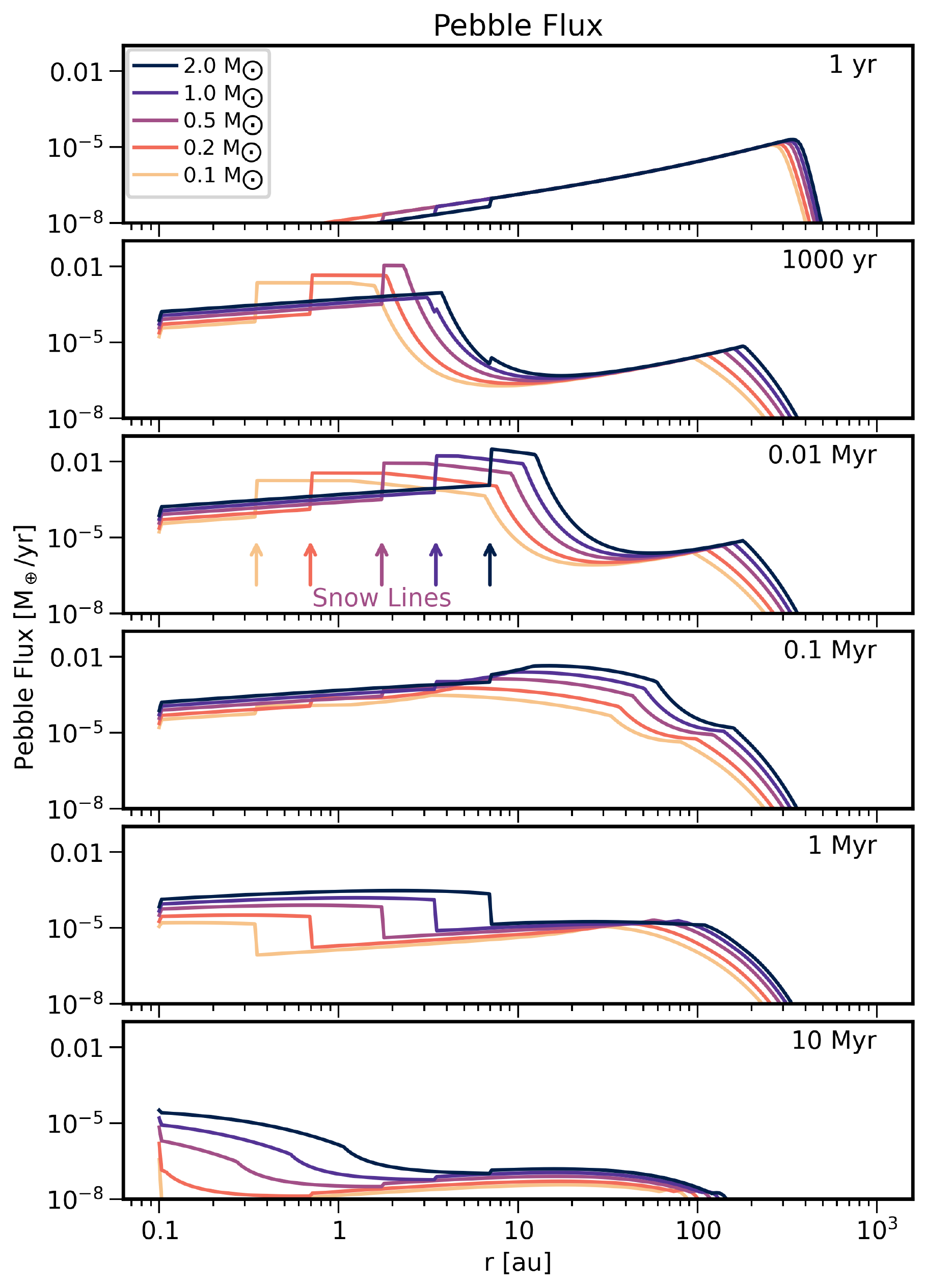}
    \else
    \includegraphics[width=\linewidth]{fig_pdf/fig2.pdf}
    \fi
    \caption{Evolution of the pebble flux in time around stars of different masses. The gas density and temperature structure are fixed in time and the location of the snowline is indicated with colored arrows.
    The pebble flux interior to the snow line is reduced at early times ($0.01$ Myr) due to increased fragmentation, but remains steady at late times ($1$ Myr) when the flux in the outer disk decays rapidly. 
    \label{f:flux}
    }
\end{figure}

\section{Pebble Accretion Model}\label{s:model}
We simulate the formation of a two-planet system through pebble accretion in an evolving protoplanetary disk.
We calculate the pebble flux at each time and location in the disks using the \texttt{pebble predictor}\footnote{\url{https://github.com/astrojoanna/pebble-predictor}} \citep{2021A&A...647A..15D}, 
and we calculate the pebble accretion efficiency onto each planetary core using the analytical fits to numerical simulations\footnote{\url{https://staff.fnwi.uva.nl/c.w.ormel/software/epsilon.tar.gz}} from \cite{2018A&A...615A.178O} and \cite{2018A&A...615A.138L}. We describe the initial conditions for the disk model and the implementation of additional mechanisms like the snow line and pebble filtering by an outer giant planet core below.

The \texttt{pebble predictor} calculates the time-dependent pebble flux based on dust growth, fragmentation, and drift time scales using a mass-weighted pebble size to describe the pebble population (see \citealt{2021A&A...647A..15D} for details). It has been benchmarked to \texttt{DustPy}\footnote{\url{https://pypi.org/project/dustpy/}} simulations that use a full dust size distribution 
and we have verified that this approximation remains valid for the lower mass stars explored in this work. Table \ref{t:disk} lists the initial disk setup and Figure \ref{f:flux} shows the pebble flux evolution for stars of different masses. 
We scale the disk gas mass and disk temperature with the stellar mass as follows:
The initial disk mass is a fixed fraction of the stellar mass, effectively a linear scaling between disk mass and stellar mass.
The disk temperature follows a square root dependence on stellar mass 
\citep[e.g.][]{2013ApJ...771..129A}. We keep all other parameters, including the disk outer radius and surface density profile, independent of stellar mass. 

\begin{table}
\begin{tabular}{l | c} 
\hline
Parameter & Value \\
\hline
$M_\star$ [$M_\odot$] & 0.1, 0.2, 0.5, \textbf{1.0}, 2.0 \\
$M_\text{disk}$ [$M_\star$] & 0.02 - 0.5 (\textbf{0.2}) \\
$R_\text{disk}$ [au] & 300 \\
$T_\text{disk}$ [$K$]& $280\, a^{-0.5}\, M_\star^{0.5}$ \\
g:d & 100 \\
$\alpha$ & $10^{-4}$ \\
\hline
$T_\text{SL}$ [K] & 150 K \\
$v_\text{frag}$ [cm/s] & 1000 \\
$v_\text{frag,SL}$ [cm/s] & \textbf{100}, 1000 \\
\hline
$M_\text{seed}$ [$M_\oplus$] & 0.01 \\
$M_\text{iso}$ [$M_\oplus$] & 40 \\
\hline
\end{tabular}
\caption{Disk Parameters. The values of the default model are highlighted in bold face. \label{t:disk}}
\end{table}

We place a water snow line in the disk at a location where the temperature reaches $150\,K$. 
This puts the snow line at 3.5 au for a solar mass stars. We remove half the solid mass interior to the snow line to account for the sublimated water mass \citep{2003ApJ...591.1220L}. Because ices are more sticky than silicate dust according to lab experiments, we reduce the fragmentation velocity interior to the snow line by a factor 10 \citep{2015ApJ...798...34G}. This leads to a smaller pebble size interior to the snow line (see also \citealt{Levison:2015jk,2015Icar..258..418M}) and as the drift velocity scales with dust size, this leads to a reduction of the pebble flux at early times ($0.01$ Myr). It also delays the drop in pebble flux at later times ($1$ Myr) when the outer disk is already drained of pebbles. 


We insert two seed planet cores of mass 0.01 $M_\oplus$ in the disk at $t=0$, one inside and one outside the snow-line. 
We place one seed at 0.3 au, representing the location of a super-Earth, and one at 5 au, representing the location of a giant planet like Jupiter. We calculate the pebble accretion efficiency onto each core using the 3D pebble accretion mode from \cite{2018A&A...615A.178O} and \cite{2018A&A...615A.138L}. At each time step, we add the accreted pebble mass to the core mass until the core reaches its pebble isolation mass \citep{2014A&A...572A.107L,2017ASSL..445..197O}, $M_\text{iso} = 40 M_\oplus \frac{M\star}{M_\odot} \left( \frac{H_p}{0.05\,a}\right)^3$, where $H_p$ is the disk pressure scale height.




Second, we make a correction such that pebbles accreted (or blocked) by the outer planet are not also accreted by the inner planet. 
The filtering factor for the inner planet is given by
\begin{equation}
f_\text{inner} = 1 - f_\text{outer}
\end{equation}
where $f_\text{outer}$ is either the fraction of accreted pebbles by the outer planet or $f_\text{outer}=1$ if the outer planet has reached pebble isolation mass.


\begin{figure}
    \centering
     \ifsubmit
    \includegraphics[width=\linewidth]{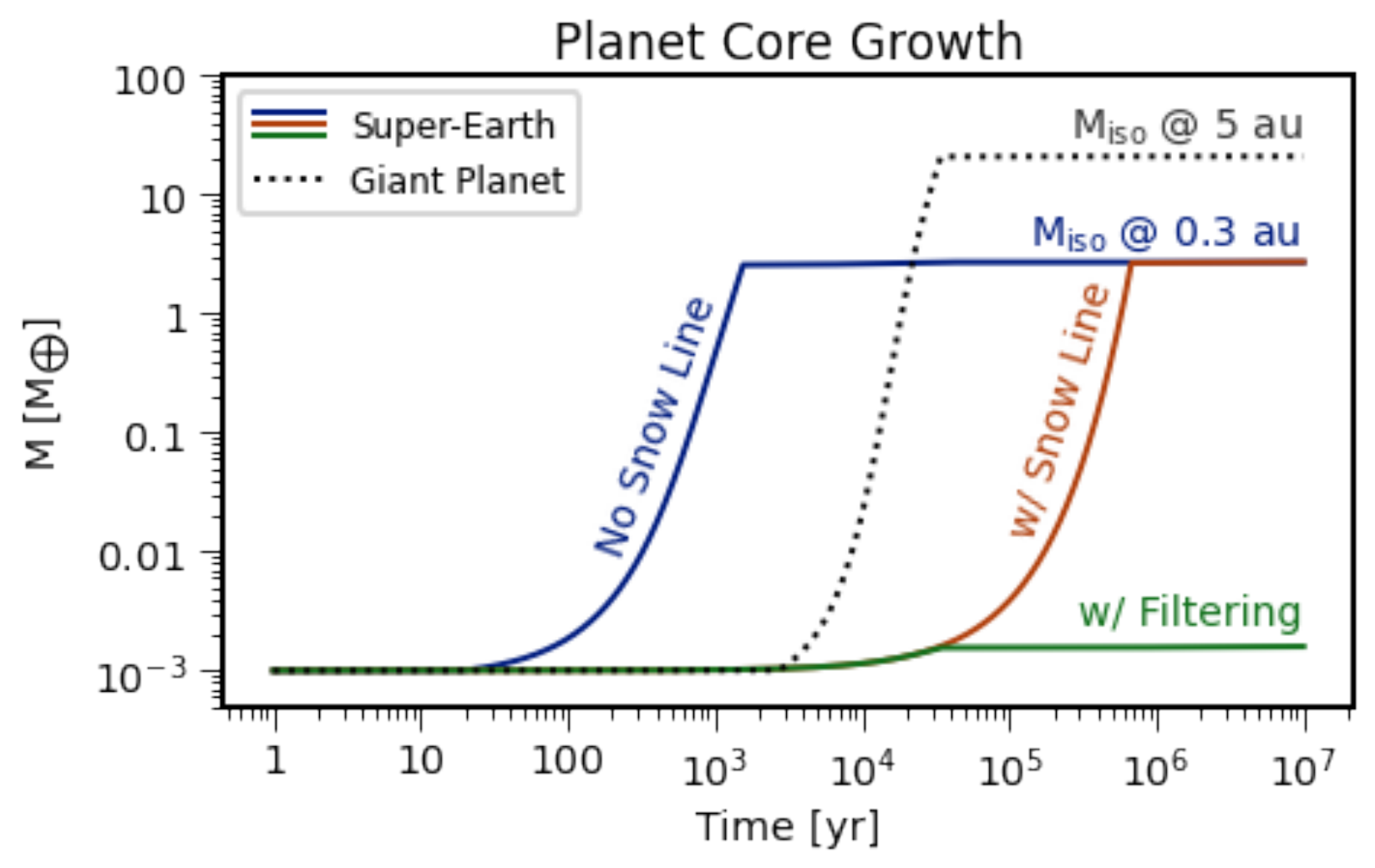}
    \else
    \includegraphics[width=\linewidth]{fig_pdf/fig3.pdf}
    \fi
    \caption{Growth of a planet core at $0.3$ au around a solar-mass star (solid lines). Without the reduction in fragmentation velocity from being inside the snow line, a super earth core (blue) grows quickly to its isolation mass before a giant planet (dotted line) can stunt its growth. With the reduced fragmentation velocity inside the snowline, the super-Earth core growth (red) is delayed until after the giant planet core forms. If filtering of pebbles by the giant planet is also taken into account (green line) a super earth does not form. 
    }
    \label{f:mass}
\end{figure}

Figure \ref{f:mass} shows the growth of such a two-planet system. 
In a model without a snow-line (blue), the inner planet grows more rapidly than the outer planet (dotted line) and reaches its isolation mass before any filtering can take place. In a model with a snowline (red), growth is more protracted because of the lower pebble flux and smaller pebble size, and a single planet at 0.3 au reaches isolation mass in just under a Myr. The planet core outside the snow line reaches its isolation mass well before that, at a few hundred thousand years. If the filtering of the giant planet core is included on the pebble flux, the inner planet ceases to grow and a super-Earth does not form (green). 

\section{Stellar Mass Dependence}\label{s:para} 
\begin{figure}
    \centering
    \ifsubmit
     \includegraphics[width=\linewidth]{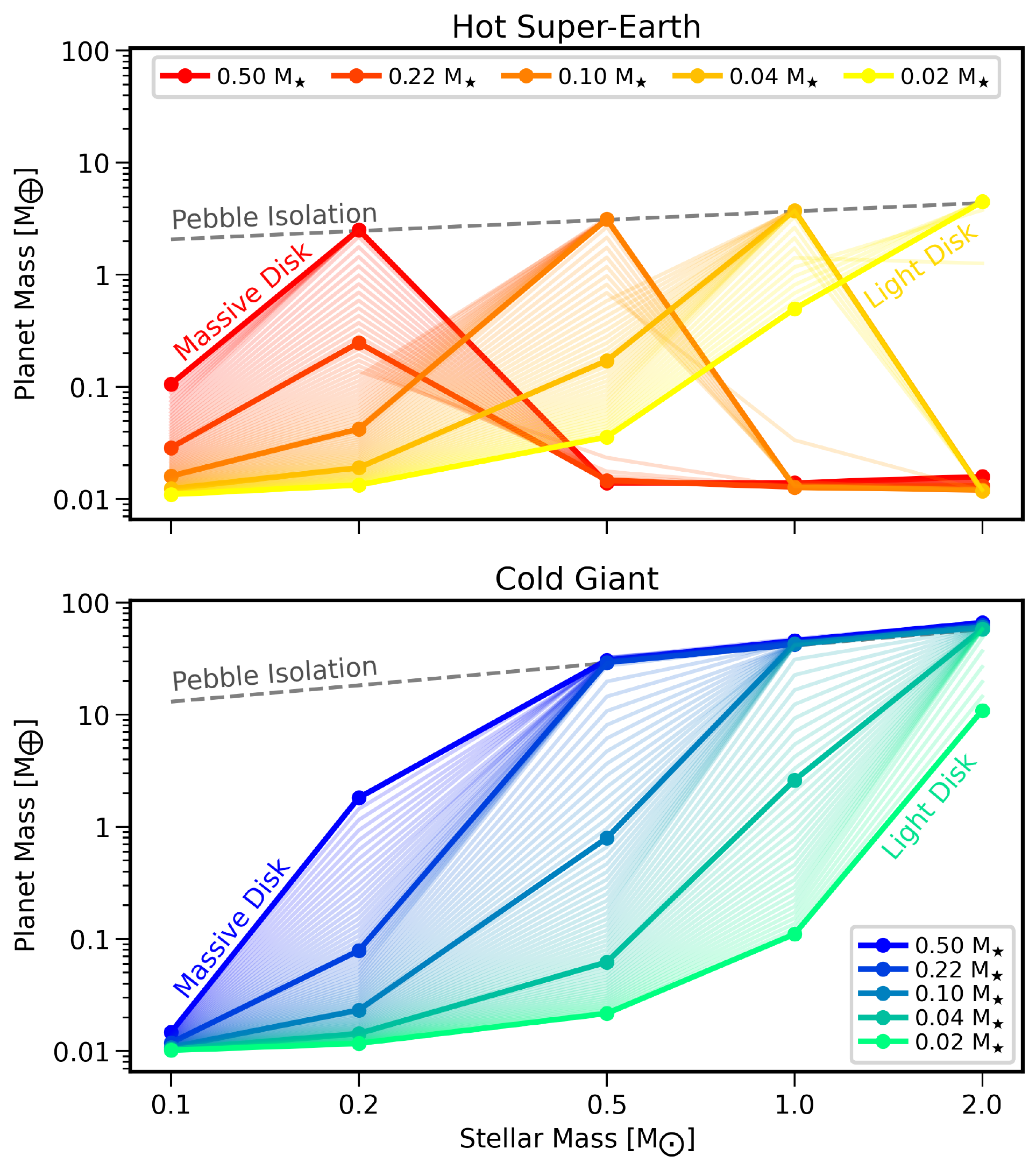}
    \else
    \includegraphics[width=\linewidth]{fig_pdf/fig4.pdf}
    \fi
    \caption{Final planet masses as a function of stellar mass for a range of initial disk masses. Disk masses are spaced equally in logarithm of disk mass with every 20th line in bold face. The dashed line denotes the isolation mass. Planets at 0.3 au (top panel) reach super-Earth mass in the highest mass disks (red) around low-mass stars. Around higher mass stars, super-Earths only form in the lower mass disks (yellow) because there giant planets do not form (bottom panel, green lines).
    }
    \label{f:mstar}
\end{figure}

Now that we have established a mechanisms through which a growing giant planet can prevent the formation of a super-Earth, we explore the dependence of this mechanism on stellar mass. We apply the pebble drift and accretion model to five different stellar masses between $0.1$ and 2 $M_\sun$. At each stellar mass, the initial disk mass is a fixed fraction of the stellar mass between $2\%$ to $50\%$. We assume a gas-to-dust ratio of $g:d=100$. 
The dispersion in disk mass is roughly representative of the range in observed millimeter fluxes of protoplanetary disks, which also span more than an order of magnitude.
Figure \ref{f:mstar} shows the resulting planet mass for a two-planet model with an inner planet growing at 0.3 au and an outer planet growing at 5 au, for a solar mass star and a disk that is $10\%$ of the stellar mass.

The outer planet follows the expected trend, growing larger in more massive disks and around more massive stars. 
Most planet cores reach isolation mass around stars of twice a solar mass, and would likely continue to grow into gas giant planets. Around stars of a solar-mass or half a solar mass, only planet cores in the most massive disks reach isolation mass and could become giant planets. 
Around the two lowest mass stars, no planet cores reach pebble isolation or become massive enough to become the cores of giant planets.  

The mass of planets in the inner disk show a more complex dependence on stellar and disk mass. Around the lowest mass stars ($0.1$-$0.2\, M_\odot$) --- and in the absence of outer giant planets --- the expected trend emerges: more massive massive disks form more massive planets. Because of the lower pebble flux, only cores in the most massive disks are able to reach isolation mass and form a super-Earth. Around solar mass and two solar mass stars, we see the exact opposite trend. In the more massive disks, the giant planet cores shut down the pebble flux to the inner disk before the cores there have time to grow, and super-Earths do not form. Only in the lowest mass disks, where giant planets do not form, are super-Earths able to form. 

The most complex dependence of super-Earth formation on disk mass is seen around stars of half a solar mass. In disks with a mass near the median ($\sim 0.1\, M_\star$, orange lines) planets reach isolation mass. Lower-mass disks produce lower-mass planets because the pebble flux is lower, while higher mass disks also produce lower mass planets but because the pebbles are filtered by the forming giant planets. Thus, there is an optimum for super-Earth formation around early M dwarfs: a high enough pebble flux to grow fast, but a low enough pebble flux to not form a giant planet.


\begin{figure}
    \centering
    \ifsubmit
    \includegraphics[width=\linewidth]{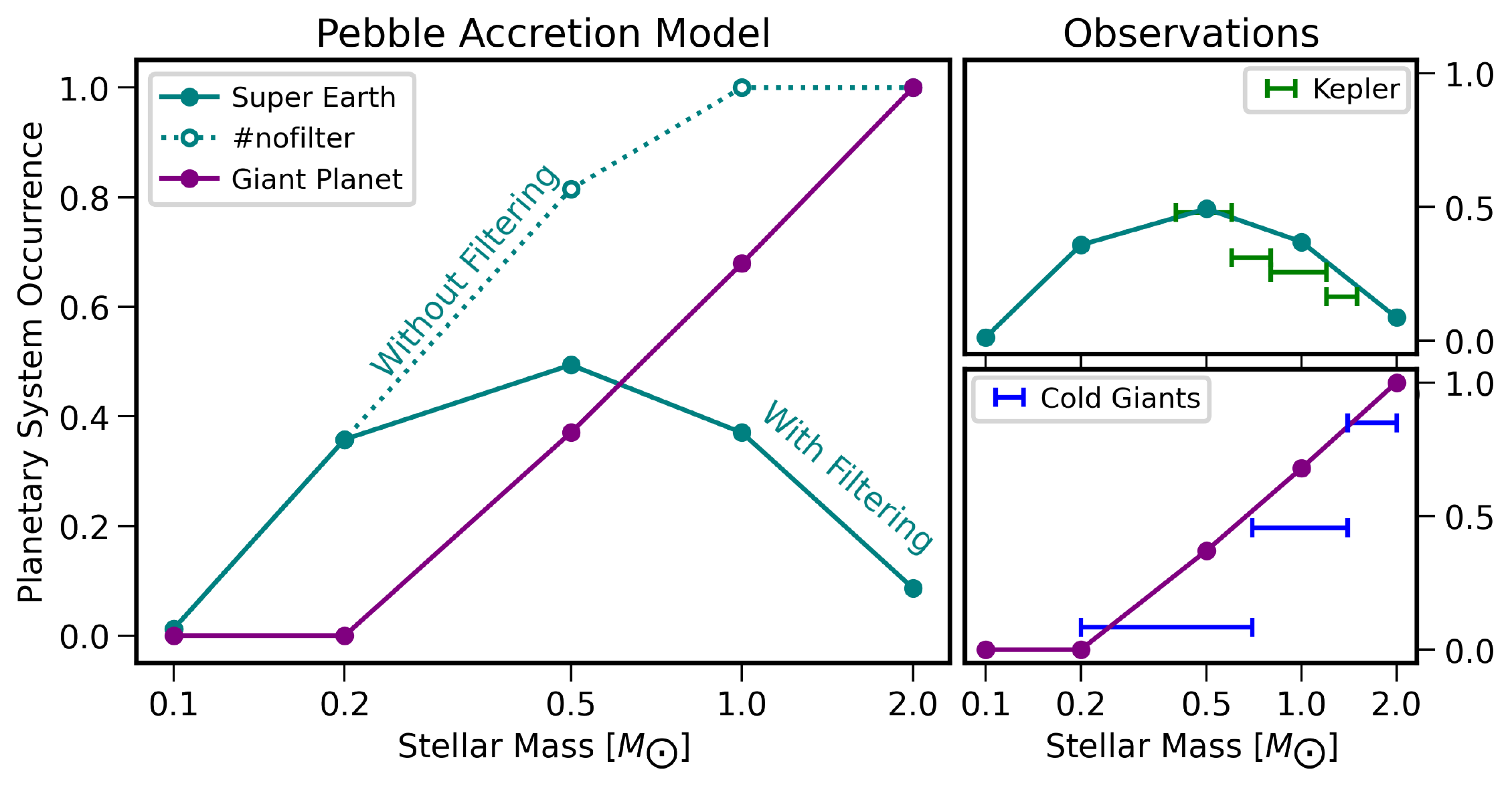}
    \else
    \includegraphics[width=\linewidth]{fig_pdf/fig5.pdf}
    \fi
    \caption{Fraction of models (from Fig. \ref{f:mstar}) that form super-Earth cores (teal) or a giant planet cores (purple). 
    For reference the fraction of models that would form super-earths in the absence of giant planet filtering is indicated with a dotted line. 
    The right panels show the observed planetary system occurrence rates from \textit{Kepler} (top) and radial velocity (bottom). 
    }
    \label{f:obs}
\end{figure}

\section{Exoplanet Occurrence Rates}\label{s:obs}
We compare the outcome of pebble accretion simulations at different stellar masses with occurrence rates of exoplanets 
from the \textit{Kepler} survey and 
from radial velocity surveys. At each stellar mass we simulate a range of disk masses from 0.02 to 0.5 times the stellar mass. We use the fraction of these simulations where the planet core reaches a certain mass as a proxy for the occurrence rate of planetary systems. Because we modeled only the growth of planetary cores through pebble accretion, and did not consider subsequent growth by 
giant impacts or growth by accretion of gas, we make a number of assumptions in this observational comparison that we outline below.

For super-Earths we compare our results with the occurrence rates of planets between 1-4 earth radius interior to 1 au for spectral types M, K, G, and F as in \cite{2015ApJ...814..130M}, but using the more recent \textit{Kepler} \texttt{DR25} dataset \citep{2018ApJS..235...38T} as described in \citep{2018AJ....156...24M}. 
We account for planet multiplicity by scaling down the occurrence rate by a factor $4.5$ to get the fraction of stars with planetary systems, following \cite{2021arXiv210712520M}. 
We consider a simulation capable of forming a detectable super-Earth system if the inner planetary core reaches $0.1 M_\oplus$, which is the embryo mass needed for formation of earth-mass planets through giant impacts \citep[e.g.][]{1998Icar..136..304C}. 

For the giant planet occurrence rate we use samples of M dwarfs, FGK stars, and retired A stars as defined in \cite{2010PASP..122..905J}. We update the planet occurrence rates for M dwarfs and FGK stars with rates from Figure 7 in \cite{2021arXiv210511584F} and with occurrence rates from \cite{2018ApJ...860..109G} for retired A stars. The stellar mass dependencies in those studies are only reported for giant planets more massive than $0.3\,M_\text{Jup}$ and $1\,M_\text{Jup}$, respectively. Therefore we apply a correction factor to scale the occurrence rates to be representative for all giant planets more massive than $0.1\,M_\text{Jup}$. We use a correction factor of $26\%/6\%\approx 4.3$ based on \cite{2019ApJ...874...81F} for the retired A stars, and a correction factor of $33\%/12\%$ based on \cite{2021arXiv210511584F}.
We consider a simulation capable of forming an observable giant planet if the outer core reaches a mass of $10 M_\oplus$, the traditional threshold for runaway gas accretion. 

Figure \ref{f:obs} shows the predicted occurrence rates of the model grid for super-Earths and giant planets as a function of stellar mass. In the right panel we compare with transit and radial velocity data. 
The simulated giant planet occurrence rate increases with stellar mass from zero at the lowest mass stars to nearly unity at twice a solar mass. This trend is qualitatively consistent with the observations and this feature is commonly reproduced in planet formation models in the core accretion framework \citep[e.g.][]{2021arXiv210504596B}. 

The simulated planet occurrence rates of super-Earths show a different trend. 
The planet occurrence rates decrease with stellar mass between $0.5$ to $2\,M_\odot$, which matches with the trend observed with \textit{Kepler} for M, K, G, and F stars. Interestingly, this trend does not continue for lower mass M dwarfs: the model predicts planet occurrence rates to decrease with stellar mass for M dwarfs of $0.1$ and $0.2\,M_\odot$. 
In the context of the pebble accretion model, it is not the planet occurrence rate of transiting planets around M dwarfs that is elevated, but the planet occurrence rate around FGK stars that is depressed. 
The dotted teal line in Figure \ref{f:obs} shows how the planet occurrence rate of super-earths around solar and super-solar mass stars would continue to increase with stellar mass in the absence of pebble-filtering giant planet cores.


\section{Conclusions and Discussion}\label{s:discussion}
We have shown using pebble drift and accretion models that forming giant planets outside the snow line can prevent the formation of close-in super-Earths. This mechanism leads to an elevated occurrence rate of transiting planets around M dwarfs as observed with the \textit{Kepler} spacecraft, 
and is also consistent with the stellar mass dependence of giant planets from radial velocity surveys. 
The presence of a snow line in the model is crucial, as it delays the growth of super-Earths and allows giant planets to reach pebble isolation mass first and cut off the flow of pebbles into the inner disk.  

A key component of this hypothesis is that giant planets suppress the formation of super-Earths. This mechanism has been previously proposed to explain the lack of super-Earths in the solar system through Jupiter \citep[e.g.][]{2015Icar..258..418M,2019A&A...627A..83L}. However, several studies have found that hot super-Earths can exist in systems with cold giant planets \citep[e.g.][]{2018ApJ...860..101Z,2019AJ....157...52B}, potentially at odds with the proposed anti-correlation. Unfortunately, sample sizes are still small and a correlation is not always found \citep{2018A&A...615A.175B}, implying that the super-Earth giant planet correlation is weak. Based on our models (Fig. \ref{f:mstar}) we predict that in systems with both super-Earths and giant planets, the planet masses should be anti-correlated: the most massive super-earths should not exists in systems that also harbor the most massive giant planets. 
A larger sample of stars monitored for both close-in super-Earths and long-period giant planets is needed, for example by combining TESS data with ground-based radial velocity surveys or with \textit{Gaia}. 


A key prediction of our model is that the occurrence rates of super-Earths decrease again for late M dwarfs. Thus, planet-hosts like Trappist-1 \citep[e.g.][]{2017Natur.542..456G} near the brown dwarf/stellar boundary may be more rare than the M dwarf planet hosts identified with Kepler, which are mostly early M dwarfs of $\approx0.4-0.5\,M_\odot$ \citep[e.g.][]{2013ApJ...767...95D}.  
\textit{Kepler} did not observe enough mid and late M dwarfs to constrain their stellar mass dependence \citep{2019AJ....158...75H}. However, intermediate results from the CARMENES radial velocity survey show changes in the planet population around low-mass ($<0.34\,M_\odot$) M dwarfs \citep{sabotta:2021ij} that should be further investigated in the context of these model predictions.



The model presented in this paper is fairly simple by design. It was chosen to illustrate a broader trend in planet populations using a small population synthesis, and not to make detailed predictions for individual exoplanet observations. To do so, it would need to be expanded with N-body interactions to model the growth phase of giant impacts and the (runaway) accretion of gas onto planet cores. Additionally, a more complete description of pebble isolation and filtering may lead to testable predictions for the correlation between super-Earths and giant planets orbiting the same stars.
Rather than expand this model further, we propose that the key mechanisms identified in this paper, e.g. the snow line, pebble isolation and filtering, be implemented in existing models. 

%
Finally, we discuss the implications for the composition of exoplanets around stars of different masses, and in particular the volatile content of M dwarf super-Earths that might be observed with JWST and other upcoming facilities.
While the pebbles accreted locally by super-Earths are expected to be volatile-poor, the water vapor released by pebbles as they drift across the snow line may be accreted onto the atmospheres of super-Earths \citep[e.g.][]{2021arXiv210902687K}. 
While giant planets can block the flow of pebbles before they reach the snow line and thus prevent the inner super-Earths from accreting water vapor \citep{Bitsch:ca} we expect this mechanism to not to operate in M dwarfs because they lack giant planets, and thus potentially be volatile rich. 

Our simulations indicate that super-Earths around M dwarfs form from a wider range of protoplanetary disk masses than super-Earths around sun-like stars, and thus may display a wider diversity in observable properties such as bulk or atmospheric composition. In addition, super-Earths form much later in the disk lifetime than giant planets in our simulations, and thus may accrete more chemically processed material. 

\begin{acknowledgments}
We would like to thank Chris Ormel and Beibei Liu for making their pebble accretion scripts available online.
We would also like to thank Lee Rosenthal and Ben Montet for sharing planet occurrence rate data.
G.D.M. acknowledges support from ANID --- Millennium Science Initiative ---  ICN12\_009. 
This material is based upon work supported by the National Aeronautics and Space Administration under Agreement No. 80NSSC21K0593 for the program ``Alien Earths''. The results reported herein benefitted from collaborations and/or information exchange within NASA's Nexus for Exoplanet System Science (NExSS) research coordination network sponsored by NASA’s Science Mission Directorate.
\end{acknowledgments}


\ifhasbib

\else	
	\bibliography{papers3,books,software}
\fi


\end{document}